\documentclass[aps,pra,superscriptaddress,twocolumn,amsmath,amssymb]{revtex4-2}
\pdfpageattr {/Group << /S /Transparency /I true /CS /DeviceRGB>>}

\usepackage{graphicx}
\usepackage{color}
\usepackage[hidelinks]{hyperref}
\usepackage{orcidlink}
\usepackage{amsmath}
\usepackage{amsthm}
\usepackage{physics}
\usepackage{amsfonts}
\usepackage{times,txfonts}
\usepackage[normalem]{ulem}

\hypersetup{colorlinks=true, linkcolor=black, citecolor=black, urlcolor=blue}

\newcommand{\fref}[1]{\text{Fig.}~\ref{#1}}

\begin{document}

\author{Christoph Hotter\,\orcidlink{0009-0003-3854-0264}}
\email[]{hottch@mit.edu}
\affiliation{Center for Hybrid Quantum Networks (Hy-Q), Niels Bohr Institute, University of Copenhagen, Jagtvej 155A, Copenhagen DK-2200, Denmark} 
\affiliation{Department of Physics, MIT-Harvard Center for Ultracold Atoms and Research Laboratory of Electronics, Massachusetts Institute of Technology, Cambridge, Massachusetts 02139, USA} 
\author{Clara Henke\,\orcidlink{0009-0004-5255-5689}} %
\affiliation{Center for Hybrid Quantum Networks (Hy-Q), Niels Bohr Institute, University of Copenhagen, Jagtvej 155A, Copenhagen DK-2200, Denmark}
\author{Cornelis Jacobus van Diepen\,\orcidlink{0000-0001-8454-2859}}
\email[]{cjvandiepen@gmail.com}
\affiliation{Center for Hybrid Quantum Networks (Hy-Q), Niels Bohr Institute, University of Copenhagen, Jagtvej 155A, Copenhagen DK-2200, Denmark}
\author{Peter Lodahl\,\orcidlink{0000-0002-9348-9591}}
\affiliation{Center for Hybrid Quantum Networks (Hy-Q), Niels Bohr Institute, University of Copenhagen, Jagtvej 155A, Copenhagen DK-2200, Denmark}
\author{Anders Søndberg Sørensen\,\orcidlink{0000-0003-1337-9163}}
\affiliation{Center for Hybrid Quantum Networks (Hy-Q), Niels Bohr Institute, University of Copenhagen, Jagtvej 155A, Copenhagen DK-2200, Denmark}

\begin{abstract}
    Quantum non-Gaussian states, which cannot be written as mixtures of Gaussian states, are necessary to achieve a quantum advantage in continuous variable systems. They represent an important benchmark for the realization of an advanced quantum light source, as they cannot be made by simple means such as displacement and squeezing. We introduce an attenuation-resistant sufficient criterion for quantum non-Gaussian states based on the second- and third-order correlation functions, $g^{(2)}$ and $g^{(3)}$. The general non-linear bound for classical mixtures of Gaussian states is $\sqrt{g^{(3)}} + 3 \sqrt{g^{(2)}} \geq 2$. Any mixture of Gaussian states must fulfill this inequality, thus, the violation of it represents a direct confirmation of quantum non-Gaussianity. We experimentally show the non-Gaussianity of the state produced by a quantum dot single-photon source, where we obtain $\sqrt{g^{(3)}} + 3 \sqrt{g^{(2)}} = 0.174 (13)$, which represents a statistical significance of more than $100$ standard deviations.  
\end{abstract}
\date{\today}
\title{Quantum Non-Gaussianity Criterion Based on Photon Correlations $g^{(2)}$ and $g^{(3)}$}

\maketitle

\emph{Introduction}---Understanding the nature of quantum states is central to the description of light. The simplest possible characterization involves only the number of photons $\langle a^\dagger a \rangle$ corresponding to the intensity of the field. Further insight into the precise nature of the quantum state can be obtained from higher-order correlation functions~\cite{Glauber1963} such as 
\begin{equation}
    g^{(2)} = \frac{\langle a^\dagger a^\dagger a a \rangle}{ \langle a^\dagger a \rangle^2 },
    \label{eq:g2}
\end{equation}
where $a$ ($a^\dagger$) describes the annihilation (creation) operator of a single harmonic oscillator mode.  
Any state of light produced by a classical source, corresponding to mixtures of coherent states $\rho_\mathrm{coh} = \sum_n p_n | \alpha_n \rangle \langle \alpha_n |$, leads to $g^{(2)} \geq 1$.  This implies that any observation of $g^{(2)} < 1$ is a direct confirmation of the non-classical nature of the source that produced it~\cite{Glauber1963, Mandel1979, Mandel1986, innocenti_coherencebased_2023}. 

Many continuous variable systems are well characterized by Gaussian states~\cite{Walls1983, weedbrook_gaussian_2012, wade_squeezing_2015, wieczorek_optimal_2015, zhang_prediction_2017, gietka_unique_2023}, which means their Wigner function is described by a Gaussian distribution of the quadratures. Examples of Gaussian states include coherent and thermal states, as well as any states produced by parametric down conversion~\cite{weedbrook_gaussian_2012, ferraro2005gaussian}. 
While Gaussian states can be non-classical under the definition above, it has been shown that they cannot achieve certain quantum advantage in e.g.\ computing, communication or sensing~\cite{lloyd_quantum_1999, bartlett_efficient_2002, braunstein_quantum_2005, niset_no-go_2009, gessner_metrological_2019, wolf_motional_2019, lachman_quantum_2022}. It is therefore highly desirable to produce non-Gaussian states, which immediately calls for methods to verify their non-Gaussian nature. 
In analogy to the above-described non-classicality bound, we here derive a bound for statistical mixtures of arbitrary Gaussian states. This means that an experimental violation provides a proof of so-called quantum non-Gaussianity (QNG)~\cite{filip_detecting_2011, jezek_experimental_2011, straka_quantum_2018, lachman_faithful_2019, lachman_quantum_2021, chabaud_certification_2021, walschaers_nongaussian_2021, lachman_quantum_2022, checchinato_losses_2024, liu_experimental_2024, kalash_certifying_2025, racz_witnessing_2025, teja_quantum_2025, bemani_heralded_2025}, which is a necessity for a quantum advantage with continuous-variable systems. 

Since $g^{(2)}$ can, in principle, take any possible value between zero and infinity for Gaussian states~\cite{Walls1983, grosse_measuring_2007}, a measure solely based on it cannot be sufficient to indicate quantum non-Gaussianity. In our approach, we therefore include the third-order correlation function 
\begin{equation}
    g^{(3)} = \frac{\langle a^\dagger a^\dagger a^\dagger a a a \rangle}{ \langle a^\dagger a \rangle^3 }
    \label{eq:g3}
\end{equation}
to obtain a quantum non-Gaussianity inequality. Such a bound, solely based on normalized correlation functions $g^{(n)}$, has the crucial advantage that it is insensitive to attenuation (besides requiring longer experimental acquisition time). Since many experiments suffer from losses and finite detection efficiency, this feature can be a major advantage in comparison to other existing non-Gaussianity criteria~\cite {filip_detecting_2011, jezek_experimental_2011, straka_quantum_2018, lachman_faithful_2019, lachman_quantum_2021, chabaud_certification_2021, walschaers_nongaussian_2021, lachman_quantum_2022, checchinato_losses_2024, liu_experimental_2024, kalash_certifying_2025, racz_witnessing_2025, lachman_hierarchies_2025, asenbeck_hierarchical_2025}.  

\emph{Gaussian states}---Any Gaussian pure state of a quantum harmonic oscillator can be expressed as a displaced squeezed state $| \xi, \alpha \rangle = D(\alpha) S(\xi) | 0 \rangle$~\cite{gerry_introductory_2023, Walls1983} with the expectation values
\begin{subequations}
\begin{eqnarray}
    & \langle a \rangle = \alpha \\
    & \langle a a \rangle = \alpha^2 - e^{i \theta} \cosh(r) \sinh(r) \\
    & \langle a^\dagger a \rangle = |\alpha|^2 + \sinh^2 (r).
\end{eqnarray}
\label{eq:squeezed_state_exp}
\end{subequations}
Here, $D(\alpha) = e^{\alpha a^\dagger - \alpha^* a}$ is the displacement operator and $S(\xi) = e^{(\xi^* a^2 - \xi {a^\dagger}^2)/2}$ the squeezing operator, with the displacement $\alpha = |\alpha| e^{i \phi}$ and the squeezing parameter $\xi = r e^{i \theta}$. 
Since a Gaussian state is fully described by its mean and variance, the above expectation values are sufficient to describe it. In other words, this also means that the second-order cumulant expansion is exact~\cite{kubo1962generalized, plankensteiner2021quantumcumulantsjl, cardin_photon-number_2024, heinzel_exploiting_2026, wick_evaluation_1950} (also known as Wick's probability theorem), which allows us to use the expressions in Eq.~\eqref{eq:squeezed_state_exp} to evaluate higher-order moments $G^{(n)} = \langle (a^\dagger)^n a^n \rangle$ yielding 
\begin{subequations}
\begin{eqnarray}
    & G^{(1)}_{G} = \langle a^\dagger a \rangle \\
    & G^{(2)}_{G} = 2 \langle a^\dagger a \rangle^2 + |\langle a a \rangle |^2 - 2 |\langle a \rangle|^4 \\
    & G^{(3)}_{G} = 6 \langle a^\dagger a \rangle^3  + 9 |\langle a a \rangle|^2 \langle a^\dagger a \rangle + 16 |\langle a \rangle|^6 \\
    & \hspace{1cm} - 18 |\langle a \rangle|^4 \langle a^\dagger a \rangle - 12 |\langle a \rangle|^2 \mathrm{Re}\{ \langle a a \rangle^*  \langle a \rangle^2 \}. \nonumber 
\end{eqnarray}
\label{eq:Gn_Gauss}
\end{subequations}
\noindent The subscript ``G" indicates the Gaussian state. See Ref.~\cite{supplement} Sec.~I for expanded expressions with the expectation values from Eq.~\eqref{eq:squeezed_state_exp}. Note that for the expressions in Eq.~\eqref{eq:Gn_Gauss} only the relative angle $\phi - \theta/2$ between the displacement $\alpha$ and the squeezing $\xi$ matters, i.e.\ we can choose $\phi = 0$ and $\alpha \geq 0$ without loss of generality, which we will use from here on. In the following, we will first derive the bound for Gaussian pure states and then show that it also holds for statistical mixtures. 

%
\begin{figure}[tb!]
    \centering
    \includegraphics[width=0.5\textwidth]{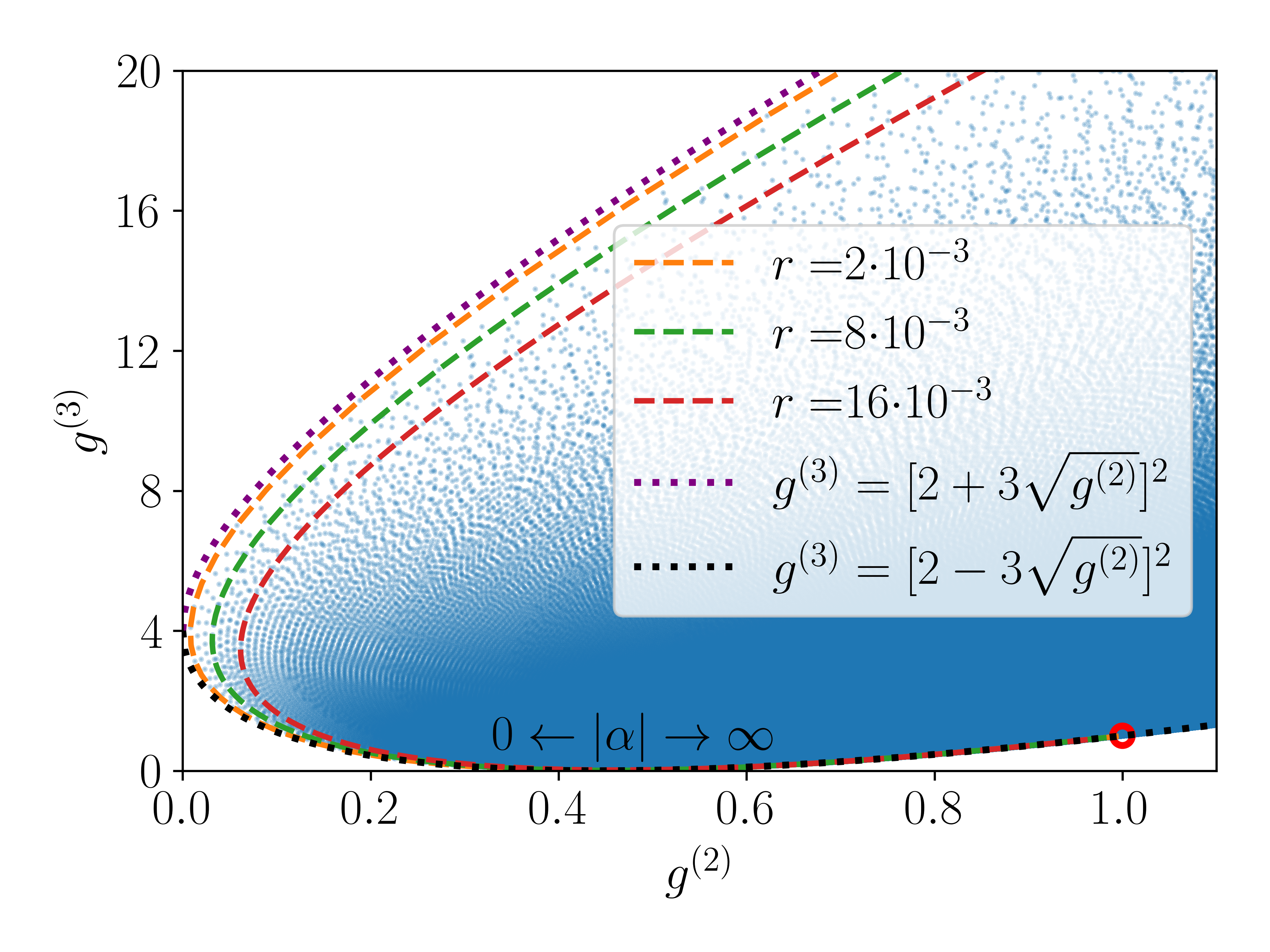}
    \caption{Second- and third-order correlation function for Gaussian states. The blue points represent $g^{(2)}$ and $g^{(3)}$ for a large set of combinations of $\alpha$, $r$ and $\theta$. The dashed lines represent fixed values of $r$ ($\theta=0$), indicating that the boundary is reached for $r \rightarrow 0$. For $|\alpha| \rightarrow \infty$ the coherent states are dominant corresponding to $g^{(2)} = g^{(3)} = 1$ (red circle). The dotted lines show the analytic upper ($g^{(3)} > 4$) and lower ($g^{(3)} < 4$) boundary for Gaussian pure states, see Eq.~\eqref{eq:g3_g2_th}. 
    For the scatter plot, we use 1000 equally distributed values in the interval $\alpha \in (0, 1]$, 501 values for $r \in [0, 1]$ and 21 values for $\theta \in [0, \pi ]$. }
    \label{fig:g2g3}
\end{figure}
\emph{Bound for Gaussian states}---Our overall goal is to derive a lower bound for $g^{(3)}$ as a function of $g^{(2)}$ for Gaussian states. We start by showing that such a bound is always given for a squeezing angle of $\theta=0$. To this end, we use a linear ansatz for the bound for statistical mixtures of Gaussian states
\begin{equation}
    g^{(3)} + \chi_2 g^{(2)} \geq \chi_1,
    \label{eq:bound_general}
\end{equation}
with so-far arbitrary real-valued factors $\chi_1$ and $\chi_2$. For Gaussian pure states, we can rewrite the inequality~\eqref{eq:bound_general} as
\begin{equation}
    G^{(3)}_G + \chi_2 G^{(2)}_G G^{(1)}_G - \chi_1 \Big[ G^{(1)}_G \Big]^3 \geq 0 .
    \label{eq:bound_general_Gn}
\end{equation}
The boundary of this inequality is reached for the minimum of the LHS with respect to the parameters of the Gaussian state, i.e.\ $\alpha$, $r$ and $\theta$. Calculating the minimum of the LHS with respect to the squeezing angle $\theta$, we find that it is always obtained for $\theta = 0$ (amplitude squeezing) if $\chi_2 \geq -3$, see \cite{supplement} Sec.~II and VI. With this minor restriction, the bound solely depends on the positive displacement $\alpha$ and the positive squeezing strength $r$. 

\begin{figure}[h]
    \centering
    \includegraphics[width=0.5\textwidth]{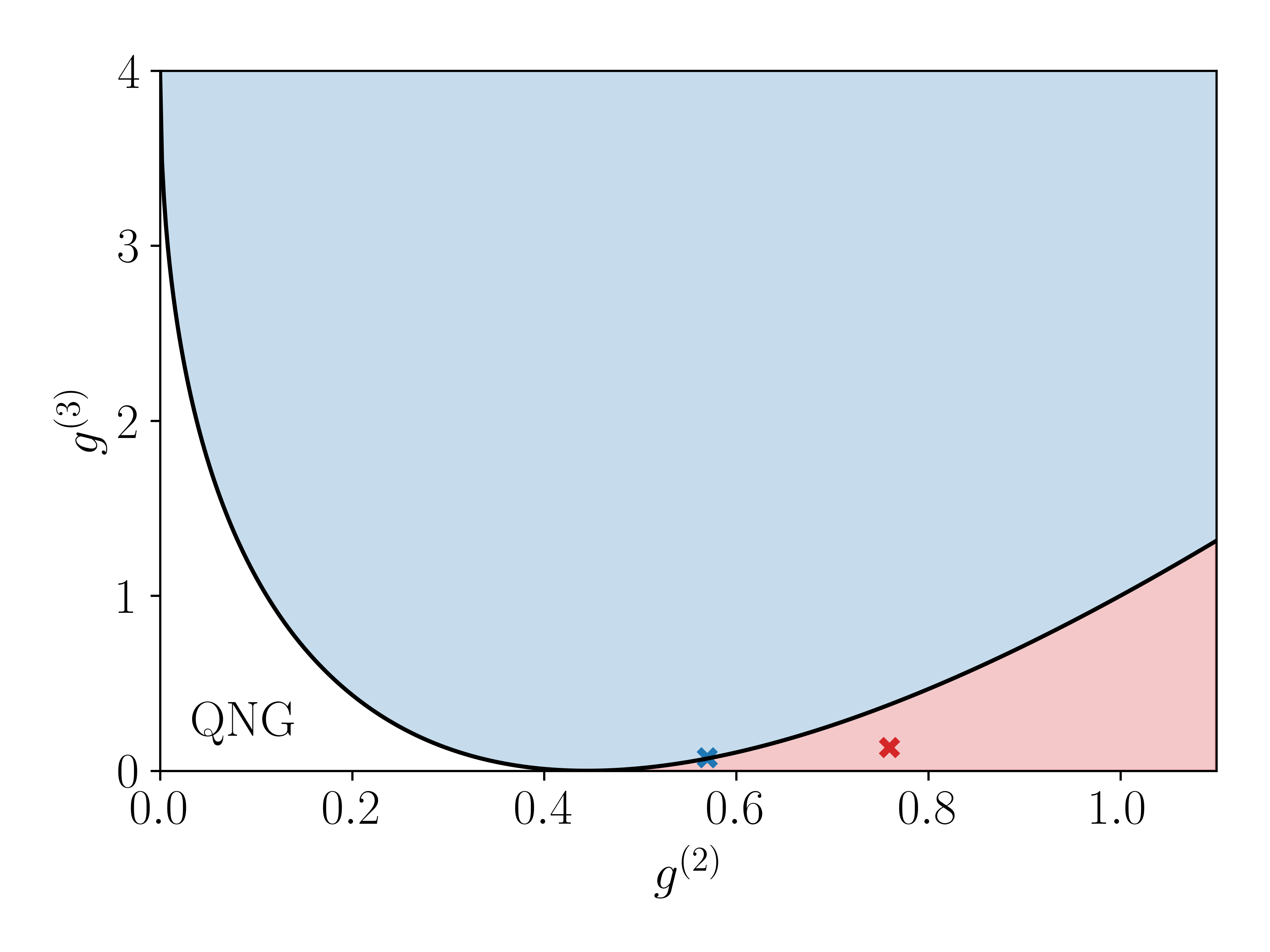}
    \caption{Non-Gaussianity bound based on $g^{(2)}$ and $g^{(3)}$. The white region cannot be reached by any incoherent superposition of Gaussian states (QNG), see Eq.~\eqref{eq:g2g3_s49}. Gaussian pure states are always above the bound in Eq.~\eqref{eq:g3_g2_bound} (black solid line, blue region). An incoherent superposition of Gaussian states can be below the bound but only for $g^{(2)} > 4/9$ (red region). 
    The blue cross close to the boundary corresponds to the Gaussian pure state $| \xi = 1/100, \alpha = 1/5 \rangle$ and the red cross to the mixed state $0.75 | \xi = 1/100, \alpha = 1/5 \rangle \langle \xi = 1/100, \alpha = 1/5 | + 0.25 | 0 \rangle \langle 0 |$.}
    \label{fig:g2g3_analytic}
\end{figure}

In order to visualize the bound for Gaussian states, we plot the values of $g^{(2)}$ and $g^{(3)}$ with the expressions in Eq.~\eqref{eq:Gn_Gauss} for a large set of values of $\alpha$, $r$ and $\theta$ in Fig.~\ref{fig:g2g3}. First of all, this scatter plot indicates that there are well-defined regions that Gaussian pure states cannot occupy and that the actual bound is non-linear. 
Furthermore, we find that states with $\theta=0$ and a fixed squeezing strength $r$ follow curves in the $(g^{(2)}, g^{(3)})$ plot, where smaller values of $r$ approach a boundary (dashed lines). 
This suggests that the boundary is obtained for $r \rightarrow 0$. However, this limit for $g^{(2)}$ and $g^{(3)}$ cannot be taken right away since that corresponds to coherent states, which have a fixed value of $g^{(n)} = 1$ for any $\alpha$ (red circle). We need to take into account that $\alpha^2$ also goes to zero for the boundary. To this end, we first expand $g^{(2)}$ and $g^{(3)}$ in a Taylor series to second order in $r$ around $r=0$, which yields $(\alpha > 0)$
\begin{align}
    g^{(2)}_{\mathrm{2nd}} = 1 - \frac{2r}{\alpha^2} + \frac{(1+2\alpha^2) r^2}{\alpha^4} + \mathcal{O}[r^3]
    \label{eq:g2_taylor}
\end{align}   
\begin{align}
    g^{(3)}_{\mathrm{2nd}} = 1 - \frac{6r}{\alpha^2} + \frac{3(3+2\alpha^2) r^2}{\alpha^4} + \mathcal{O}[r^3].
    \label{eq:g3_taylor}
\end{align}  
Neglecting the higher order terms $\mathcal{O}[r^3]$ and rearranging Eq.~\eqref{eq:g2_taylor} to find $\alpha^2$ leads to the two solutions 
\begin{align}
    \alpha^2_{\mathrm{b},\pm} = \frac{r^2-r \pm \sqrt{g^{(2)} r^2 - 2r^3 + r^4}}{g^{(2)} - 1}.
    \label{eq:a2_th}
\end{align}   
Inserting this expression in the Taylor expanded third-order correlation function of a Gaussian state $g^{(3)}_\mathrm{2nd}$ [Eq.~\eqref{eq:g3_taylor}] and taking the limit $r \rightarrow 0$, we obtain 
\begin{align}
    g^{(3)}_{\mathrm{b},\pm} = \Big( 2 \pm 3 \sqrt{g^{(2)}} \Big)^2.
    \label{eq:g3_g2_th}
\end{align}   
This corresponds to the upper and lower threshold curves in Fig.~\ref{fig:g2g3} (dotted lines). While the assumption $\theta = 0$ with the requirement $\chi_2 \geq -3$ is not fulfilled for the upper threshold curve, the numerical simulations in Fig.~\ref{fig:g2g3} still suggest that it is a valid bound for Gaussian pure states. However, statistical mixtures of Gaussian states can reach the region above the upper threshold curve, which we verified numerically. Regardless of this, we will focus only on the lower bound $g^{(3)}_{\mathrm{b},-}$ for the remainder of this paper, where the restriction $\chi_2 \geq -3$ is always fulfilled, see Sec.~VI of the supplementary material~\cite{supplement}. From the above equation, we see that the lower bound reaches $g^{(3)} = 0$ for  $g^{(2)} = 4/9$ and $g^{(2)} = 0$ at $g^{(3)} = 4$, see also Fig.~\ref{fig:g2g3_analytic}. 

\emph{Proof of the inequality for Gaussian pure states}---The derivation in the previous part only serves as an indicator for the boundary. Here, we prove that the expression for $g^{(3)}_{\mathrm{b},-}$ is indeed a threshold for Gaussian pure states, i.e.\ that they fulfill
\begin{align}
    g^{(3)} - \Big( 2 - 3 \sqrt{g^{(2)}} \Big)^2 \geq 0.
    \label{eq:g3_g2_bound}
\end{align} 
Expanding the square, multiplying the inequality by $\Big[ G^{(1)} \Big]^3$ and rearranging it, we obtain 
\begin{align}
    12 \sqrt{G^{(2)}} \Big[ G^{(1)} \Big]^2 \geq 9 G^{(2)} G^{(1)} + 4 \Big[ G^{1} \Big]^3 - G^{(3)}.
    \label{eq:G_bound}
\end{align} 
Now we consider two cases. First, if the RHS is negative, the inequality is trivially fulfilled since $G^{(1)}$ and $G^{(2)}$ are positive. Second, for a positive RHS, we are allowed to square the inequality. Inserting the expressions for $G^{(n)}_G$ from Eq.~\eqref{eq:Gn_Gauss} with the expectation values for Gaussian states in Eq.~\eqref{eq:squeezed_state_exp}, utilizing $\theta = 0$ according to the derivation above, dividing the inequality by $4 \sinh^4(r)$ (for readability) and rearranging leads to 
\begin{widetext}
\begin{equation}
\begin{aligned}
&36 \alpha^6 \cosh^2(r) - 36 \alpha^6 \cosh(r) \sinh(r) + 12 \alpha^6 \sinh^2(r) + 54 \alpha^4 \cosh^2(r) \sinh^2(r) - 48 \alpha^4 \cosh(r) \sinh^3(r) + \\ 
&+ 21 \alpha^4 \sinh^4(r) + 36 \alpha^2 \cosh^2(r) \sinh^4(r) - 
 18 \alpha^2 \cosh(r) \sinh^5(r) + 12 \alpha^2 \sinh^6(r) + 9 \cosh^2(r) \sinh^6(r) + 
 2 \sinh^8(r) \geq 0.
 \end{aligned}
    \label{eq:G_bound_Gauss_3}
\end{equation} 
\end{widetext}
Since $\cosh(r) \geq \sinh(r)$, we see that for every negative term, the positive term in front of it is larger, hence the inequality holds for Gaussian pure states. 

\emph{Proof of the inequality for Gaussian mixed states}---Above we have seen that the bound~\eqref{eq:g3_g2_bound} holds for Gaussian pure states. In this part, we generalize the result to statistical mixtures of Gaussian states $\rho_\mathrm{G} = \sum_i p_i | \xi_i, \alpha_i \rangle \langle \xi_i, \alpha_i |$ with $\sum_i p_i = 1$. 
Note that this class of states also includes thermal states, since they have a positive Glauber-Sudarshan P-representation~\cite{Glauber1963,sudarshan_equivalence_1963}, as well as displaced squeezed thermal states. 
We first emphasize that the inequality can only hold for $g^{(2)} < 4/9$, which means that the lower left corner in Fig.~\ref{fig:g2g3} and Fig.~\ref{fig:g2g3_analytic} cannot be populated with statistical mixtures of Gaussian states. For the region with $g^{(2)} > 4/9$, statistical mixtures can be below the bound, as will be shown later.  

We start by rearranging inequality~\eqref{eq:g3_g2_bound} to 
\begin{align}
    g^{(3)} \geq \Big( 2 - 3 \sqrt{g^{(2)}} \Big)^2.
    \label{eq:g3_g2_bound_mixed}
\end{align} 
We are allowed to take the square root on both sides only if the expression $2 - 3 \sqrt{g^{(2)}}$ is positive, which corresponds to $g^{(2)} < 4/9$. Taking the square root and multiplying the inequality by $\sqrt{G^{(1)}}^3$ we obtain 
\begin{align}
    \sqrt{ G^{(3)} } + 3 \sqrt{G^{(2)}} \sqrt{G^{(1)}} - 2 {\sqrt{G^{(1)}}}^3 \geq 0.
    \label{eq:g3_g2_bound_mixed2}
\end{align} 
This holds for any Gaussian pure states, which means it also holds for a weighted sum 
\begin{align}
    \sum_i p_i \Bigg[ \sqrt{ G_i^{(3)} } + 3 \sqrt{G_i^{(2)}} \sqrt{G_i^{(1)}} - 2 {\sqrt{G_i^{(1)}}}^3 \Bigg] \geq 0, 
    \label{eq:g3_g2_bound_mixed3}
\end{align} 
where $\sum_i p_i = 1$. To prove that the inequality~\eqref{eq:g3_g2_bound_mixed2} also holds for Gaussian mixed states, we need to show that it also holds for $G^{(n)} \rightarrow \sum_i p_i G_i^{(n)}$, i.e.\ that
\begin{align}
    \sqrt{ \sum_i p_i G_i^{(3)} } + 3 \sqrt{\sum_i p_i G_i^{(2)}} \sqrt{\sum_i p_i G_i^{(1)}} - 2 {\sqrt{\sum_i p_i G_i^{(1)}}}^3 \geq 0.
    \label{eq:g3_g2_bound_mixed_weighted}
\end{align} 
For the first term, we use Jensen's inequality for concave functions~\cite{Jensen1906}. The second term is approximated by the Cauchy-Schwarz inequality~\cite{cauchy1821formules}, and for the last term, we use Jensen's inequality for convex functions~\cite{Jensen1906}. 
Using all of these estimations~\cite{steele2004cauchy, supplement}, we find that the LHS is lower bounded by the LHS of the inequality~\eqref{eq:g3_g2_bound_mixed3} for weighted pure states. This proves that it also holds for statistical mixtures of Gaussian states if $g^{(2)} < 4/9$. 
This means, measuring a combination of $g^{(2)}$ and $g^{(3)}$ with 
\begin{align}
 \sqrt{g^{(3)}} + 3 \sqrt{g^{(2)}} < 2 
 \label{eq:g2g3_s49}
\end{align}
is an unambiguous proof of quantum non-Gaussianity for single-mode fields. In the supplementary material~\cite{supplement} Sec.~V, we show that the inequality also holds for Gaussian multi-mode fields. 

In the previous part, we have shown that mixtures of Gaussian states cannot reach the area below the bound~\eqref{eq:g3_g2_bound} for $g^{(2)} < 4/9$. We also argued that we do not expect the bound to hold for mixtures of Gaussian states if $g^{(2)} > 4/9$. This can be shown with a counterexample by mixing the vacuum state with a Gaussian pure state. For example, the incoherent superposition with $75\%$ of the Gaussian pure state $| \xi = 1/100, \alpha = 1/5 \rangle$ (blue cross in Fig.~\ref{fig:g2g3_analytic}) and $25\%$ vacuum $|0\rangle$ is below the boundary (red cross in Fig.~\ref{fig:g2g3_analytic}). Also, any incoherent superposition of a coherent state $| \alpha \neq 0 \rangle$ and more than $50 \%$ vacuum is below the bound. 
Furthermore, since there are Gaussian states with finite $g^{(2)} = 4/9$ and arbitrarily small $g^{(3)}$, any point with $g^{(2)} > 4/9$ can be reached by mixing vacuum to it. The same argument holds for states above $g^{(3)} > 4$, since there are Gaussian states with finite $g^{(3)} = 4$ and arbitrarily small $g^{(2)}$. 
This means the measure only strictly holds for $g^{(2)} < 4/9$. Nevertheless, the bound~\eqref{eq:g3_g2_bound} is valid for Gaussian pure states, which can be useful for theoretical considerations and circumstances where pure states can be ensured. 

\begin{figure}
    \centering
    \includegraphics[width=1\linewidth]{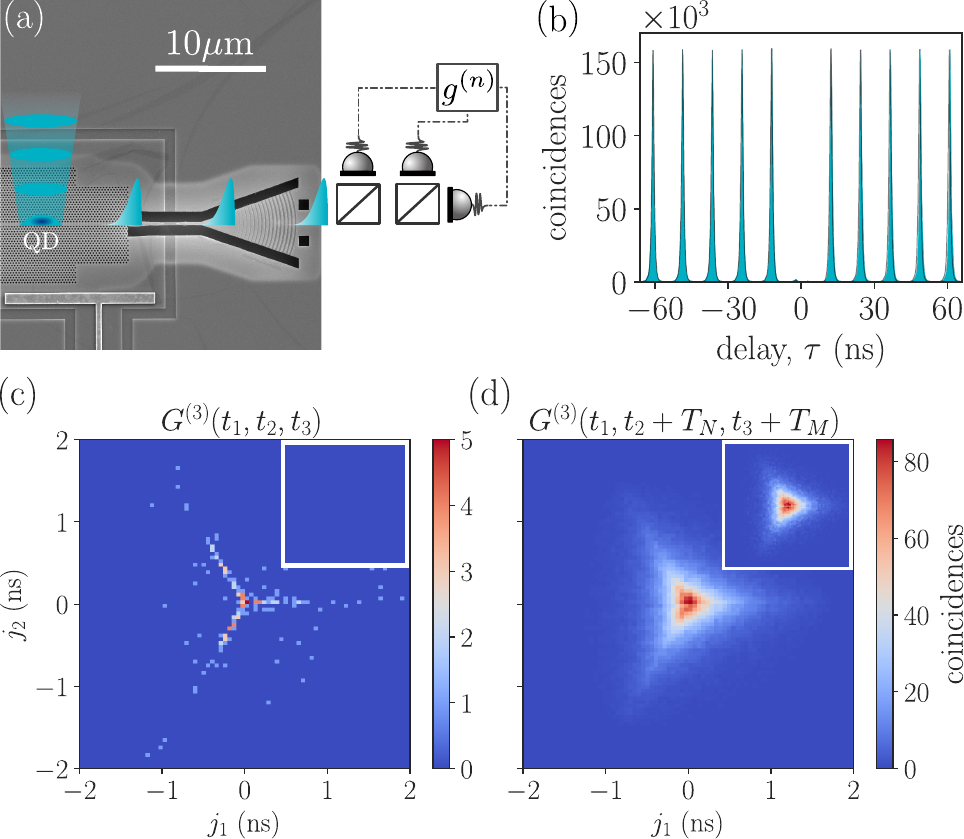}
    \caption{Experimental validation with a single-photon source. (a) Scanning electron microscopy image of the device. The quantum dot, pulsed laser excitation and emitted photons are schematically indicated. The emitted photons are routed to three single-photon detectors via a pair of beam splitters to measure coincidence counts and extract $g^{(n)}$.
    (b) Measured two-fold coincidence counts to extract the second-order intensity correlations, $g^{(2)}$. The coincidences at zero delay are strongly suppressed in comparison to the peaks with detections from separate excitation pulses. (c), (d) Measured three-fold coincidences for (c) the same excitation pulse and (d) three separate pulses after transformation and projection onto Jacobi coordinates $j_1 = (2t_1 - t_2 - t_3)/\sqrt{6}$ and $j_2 = (t_2-t_3)/\sqrt{2}$ with $t_i$ the time at detector $i$. The main panels show data with intentionally increased laser leakage, while the insets show the data for optimized conditions.} 
    \label{fig:3}
\end{figure}

\emph{Experimental validation}---Finally, we utilize the above-derived criterion to certify the non-Gaussianity for a well-established single-photon source. The source, depicted in \fref{fig:3}(a), is based on an InAs quantum dot (QD) embedded in a photonic crystal waveguide~\cite{Lodahl2015, Henke2025}. The QD is electrically tuned using a p-i-n diode structure to operate in the neutral exciton regime. An excitation is induced transversely from free space by a pulsed resonant laser with the power calibrated to a $\pi$ pulse. The emitted photons are coupled out of the device with a grating coupler connected to the end of the waveguide and collected into a fiber. Two-fold and three-fold coincidences~\cite{stiesdal2018observation, liang_observation_2018} are measured by routing the outcoupled light via a pair of beam splitters to three superconducting nanowire single-photon detectors. 

The measured two-fold coincidence counts are shown in \fref{fig:3}(b). At a delay $\tau=0$ ns, the peak is strongly suppressed compared to the side peaks, representative of a high-purity single-photon source. To extract a value for $g^{(2)}$, the integrated coincidence peak in time-resolved $g^{(2)}(t_1,t_2)$ around $\tau=t_1-t_2=0$ is normalized to the peak at $\tau=500 T$, with $T=12.15$ ns being the laser pulse repetition period. The normalization is performed with a further delayed peak to mitigate the potential effect of blinking, which is nevertheless found to be negligible for this source. 

The three-fold coincidences are shown in \fref{fig:3}(c) and (d) for detections from the same excitation pulse and from three separate pulses, respectively. As a check for our setup and analysis, an experimental run was performed with intentionally increased laser background by misalignment of the optical excitation path and removing a spectral filter in the collection. The observed three-fold rotational symmetry ($120^\circ$ in Jacobi coordinates) is a consequence of the equivalence of the three detector times and the asymmetric temporal profile of the photons emitted by the QD. The narrow feature in \fref{fig:3}(c) is due to two detections corresponding to photons from the laser background, which is temporally narrow (corresponding to e.g.\ $j_2=0$ if these clicks occur in detectors 1 and 2), while the third detection is from the QD emission, which has an exponential decay profile. Under optimized conditions, see insets, zero three-fold coincidences are detected in the emission for the same laser pulse while $6.8 \cdot 10^3$ events are recorded for photons in three separate pulses. Note that these are the statistics accumulated during a measurement time of effectively one hour, which corresponds to $3\cdot 10^{11}$ laser pulses. Further information on the three-fold coincidence measurements is provided in Sec.~VII of Ref.~\cite{supplement}.

The extracted correlation values are $g^{(2)} = 0.00334(4)$, and $g^{(3)} = 0$ with the upper bound at one $\sigma$-level of $1.7 \cdot 10^{-4}$. We use an integration window of $3.2$ ns for each time coordinate, and the errors are based on Poisson statistics. 
This is well into the bottom left region in \fref{fig:g2g3_analytic}, and the criterion yields $\sqrt{g^{(3)}} + 3 \sqrt{g^{(2)}} = 0.174(13) < 2$, thus it is clearly satisfied. 
This shows that the QD single-photon source produces non-Gaussian states away from the bound by more than $100\sigma$. 
In addition, we perform a p-value test under the hypothesis that the measured state is Gaussian, see Sec.~VIII of the supplementary material~\cite{supplement}. 
We calculate an extremely small p-value $p=4\cdot 10^{-4793}$, allowing us to reject the null-hypothesis with very high confidence. 

We emphasize here that the conventional analysis of Hanbury Brown–Twiss type measurement \cite{Brown1954, Brown1956}, where $\langle a^\dagger a\rangle$ is approximated by the detection probability, may deviate from Glauber's definition of $g^{(n)}$~\cite{Glauber1963}. The discrepancy appears in the case of high collection efficiencies for photon counters that only measure the presence of photons and not their number. 
In our experimental setup, the efficiency is sufficiently low to match the Glauber definition, which is the relevant quantity for our bound. For high detection efficiencies, a photon number resolving detection or other means are required~\cite{lachman_quantum_2022}. 

\emph{Conclusion}---We have derived a sufficient criterion to verify quantum non-Gaussianity based on the second- and third-order correlation functions $g^{(2)}$ and $g^{(3)}$. To this end, we first calculated the non-linear bound $g^{(3)} \geq (2 - 3 \sqrt{g^{(2)}})^2$ for Gaussian pure states. Building on this, we showed that satisfying the inequality $\sqrt{g^{(3)}} + 3 \sqrt{g^{(2)}} < 2$ is an unambiguous proof of quantum non-Gaussianity. Since the criterion is solely based on $g^{(2)}$ and $g^{(3)}$, it is fundamentally resistant to losses and finite detection efficiencies, which are typically limiting factors for verifying quantum non-Gaussianity. 
Finally, utilizing a quantum dot in a photonic crystal waveguide as a single-photon source, we were able to fulfill this condition with doubtless statistical significance of more than $100 \sigma$. 

The non-classicality criterion of light $g^{(2)} < 1$ was an important landmark in quantum optics. Including $g^{(3)}$ to characterize quantum states is a natural extension, which can serve as a benchmark for more advanced quantum light sources. 

Although our criterion is resistant to loss, we want to emphasize that dark counts and mixing with coherent or thermal states can be harmful due to the corresponding increase of $g^{(2)}$ and $g^{(3)}$. Furthermore, since a Fock state $|n\rangle$ has a $g^{(2)}$ of $1-1/n$ and our bound requires $g^{(2)} < 4/9$, the quantum non-Gaussianity for states with $n \geq 2$ cannot be detected, and multi-photon contributions can be harmful for our criterion, see Sec.~IX of Ref.~\cite{supplement}. 
Future works may include higher-order correlations such as $g^{(4)}$ to extend the number of non-Gaussian states that can be verified. 
An additional criterion based on the mean photon number $\langle a^\dagger a \rangle$ and $g^{(2)}$ is shown in Sec.~X of the supplementary material~\cite{supplement}. 

\emph{Acknowledgment}---C.H.\ and A.S.\ acknowledge discussions with R.\ Filip, who independently derived the quantum non-Gaussian criterion for correlation functions presented in the supplementary material~\cite{kalash_certifying_2025, racz_witnessing_2025}; The authors acknowledge discussions with K.\ Mølmer, J.\ Alba, B.\ Tissot, N.\ Kaufmann, J.\ Pinske, O.\ Sandberg and L.\ Hansen. A.S.\ acknowledges discussions with T.\ Petersen and A.\ M.\ Søndberg Sørensen. The authors gratefully acknowledge financial support from Danmarks Grundforskningsfond (Grant No. DNRF139, Hy-Q Center for Hybrid Quantum Networks). C.H.\ was supported by the Carlsberg Foundation through the ``Semper Ardens" Research Project QCooL. This research was funded in whole or in part by the Austrian Science Fund (FWF) 10.55776/J4865. For the purpose of open access, the author has applied a CC BY public copyright license to any Author Accepted Manuscript (AAM) version arising from this submission. The data presented in this article is available from Ref.~\cite{zenodo}. 

\bibliography{references}

\clearpage
\newpage

\onecolumngrid

\setcounter{equation}{0}
\setcounter{figure}{0}
\renewcommand{\theequation}{S\arabic{equation}}
\renewcommand{\thefigure}{S\arabic{figure}}

\section*{Supplemental Material: Quantum Non-Gaussianity Criterion Based on Photon Correlations $g^{(2)}$ and $g^{(3)}$}

\section{Expanded expressions of the correlation functions for Gaussian states $G^{(n)}_{G}$}
In this section, we show the expressions for $G^{(1)}_\mathrm{G}$, $G^{(2)}_\mathrm{G}$ and $G^{(3)}_\mathrm{G}$ for displaced squeezed states (Gaussian states), i.e.\ we insert Eq.~(3) into Eq.~(4) with $\phi = 0$.
\begin{align*}
    G^{(1)}_{G} &= \langle a^\dagger a \rangle = |\alpha|^2 + \sinh^2(r) \nonumber \\ 
    G^{(2)}_{G} &= 2 \langle a^\dagger a \rangle^2 + |\langle a a \rangle |^2 - 2 |\langle a \rangle |^4 = \cosh^2(r) \sinh^2(r) + 2 \sinh^4(r) + 4 |\alpha|^2 \sinh^2(r) - 2 |\alpha|^2 \cosh(r) \sinh(r) \cos(\theta) + |\alpha|^4  \nonumber \\ 
    G^{(3)}_{G} &= 6 \langle a^\dagger a \rangle^3  + 9 |\langle a a \rangle|^2 \langle a^\dagger a \rangle + 16 |\langle a \rangle|^6 - 18 |\langle a \rangle|^4 \langle a^\dagger a \rangle - 12 |\langle a \rangle|^2 \mathrm{Re}\{ \langle a a \rangle  \langle a \rangle^2 \} \\ \nonumber
    &= 6 \sinh^6(r) + 9 \sinh^4(r) \cosh^2(r) + |\alpha|^2 \left[ 18 \sinh^4(r) + 9 \sinh^2(r) \cosh^2(r) - 18 \sinh^3(r) \cosh(r) \cos(\theta) \right] \\
    &+ |\alpha|^4 \left[ 9 \sinh^2(r) - 6 \cosh(r) \sinh(r) \cos(\theta) \right] + |\alpha|^6   \nonumber
\end{align*}

\section{Minimum with respect to $\theta$}
In this section, we show that the bound is always given for amplitude squeezed states, i.e.\ for $\theta = 0$ (with $\phi = 0$). This means that for $G_\mathrm{G}^{(3)} + \chi_2 G_\mathrm{G}^{(2)} G_\mathrm{G}^{(1)} + \chi_1 [G_\mathrm{G}^{(1)}]^3$ the minimum is always obtained for $\theta = 0$ if $\chi_2 \geq -3$, which can be seen with the following:
\begin{alignat*}{2}
    & \partial_\theta G_\mathrm{G}^{(1)} = 0 \\
    &\partial_\theta \Big[ G_\mathrm{G}^{(3)} + \chi_2 G_\mathrm{G}^{(2)} G_\mathrm{G}^{(1)} \big] &&= \Big[ 18 |\alpha|^2 \sinh^3(r) \cosh(r) + 6 |\alpha|^4 \cosh(r) \sinh(r) \Big] \sin(\theta) \\
    & &&+ \chi_2 \Big[ 2|\alpha|^2 \sinh^3(r) \cosh(r) + 2 |\alpha|^4 \cosh(r) \sinh(r) \Big] \sin(\theta) \\
    &\partial^2_\theta \Big[ G_\mathrm{G}^{(3)} + \chi_2 G_\mathrm{G}^{(2)} G_\mathrm{G}^{(1)} \big] &&= \Big[ 18 |\alpha|^2 \sinh^3(r) \cosh(r) + 6 |\alpha|^4 \cosh(r) \sinh(r) \Big] \cos(\theta) \\
    & &&+ \chi_2 \Big[ 2 |\alpha|^2 \sinh^3(r) \cosh(r) + 2 |\alpha|^4 \cosh(r) \sinh(r) \Big] \cos(\theta)
\end{alignat*}
For $\theta = 0$ and $\chi_2 \geq -3$, the second derivative is always positive, i.e.\ it corresponds to a minimum. The restriction $\chi_2 \geq -3$ corresponds to a tangent line to the bound with a slope of $k \leq 3$. For the derived lower bound~(11), this is always fulfilled. In Sec.~\ref{sec:liner_bound}, we show the expression for the slope and visualize it. 

\section{Jensen's and Cauchy-Schwarz inequalities for the mixed states proof}
To prove the inequality~(17) for Gaussian mixed states, we use the following inequalities. Jensen's inequality for concave functions~\cite{Jensen1906}: 
\begin{align*}
 \sqrt{ \sum_i p_i G_i^{(3)} } \geq \sum_i p_i \sqrt{ G_i^{(3)} } 
\end{align*}
Cauchy-Schwarz inequality~\cite{cauchy1821formules}: 
\begin{align*}
    \sqrt{\sum_i p_i G_i^{(2)}} \sqrt{\sum_i p_i G_i^{(1)}} &\geq \sum_i \sqrt{p_i G_i^{(2)}} \sqrt{p_i G_i^{(1)}} = \sum_i p_i  \sqrt{G_i^{(2)}} \sqrt{G_i^{(1)}} 
\end{align*}
Jensen's inequality for convex functions~\cite{Jensen1906}:
\begin{align*}
 \left[ \sum_i p_i G_i^{(1)} \right]^{(3/2)} \leq \sum_i  p_i  \Big[G_i^{(1)} \Big]^{(3/2)}. 
\end{align*}
Using these results, we can see that
\begin{align}
    \sqrt{ \sum_i p_i G_i^{(3)} } + 3 \sqrt{\sum_i p_i G_i^{(2)}} \sqrt{\sum_i p_i G_i^{(1)}} - 2 {\sqrt{\sum_i p_i G_i^{(1)}}}^3 \geq \sum_i p_i \Bigg[ \sqrt{ G_i^{(3)} } + 3 \sqrt{G_i^{(2)}} \sqrt{G_i^{(1)}} - 2 {\sqrt{G_i^{(1)}}}^3 \Bigg] \geq 0, 
    \label{eq:g3_g2_bound_mixed_weighted_supp}
\end{align} 
which proves the inequality for Gaussian mixed states. 

\section{Useful expressions for Gaussian states}
Here, we show further useful expressions for displaced squeezed states. 
\begin{align*}
    \Big[ G_\mathrm{G}^{(1)} \Big]^3 &= |\alpha|^6 + 3 |\alpha|^4 \sinh^2(r) + 3 |\alpha|^2 \sinh^4(r) + \sinh^6(r) \\
    G_\mathrm{G}^{(2)} G_\mathrm{G}^{(1)} &= |\alpha|^6 + 5 |\alpha|^4 \sinh^2(r) + 2 \sinh^6(r) + \sinh^4(r) \cosh^2(r) + 6 |\alpha|^2 \sinh^4(r) + |\alpha|^2 \sinh^2(r) \cosh^2(r) \\
    &- 2 |\alpha|^4 \sinh(r) \cosh(r) \cos(\theta) - 2 |\alpha|^2 \sinh^3(r) \cosh(r) \cos(\theta) \\
    |\langle a a \rangle|^2 &= |\alpha|^4 - 2 |\alpha|^2 \cosh(r) \sinh(r) \cos(\theta)  + \cosh^2(r) \sinh^2(r) \\
    \langle a^\dagger a \rangle^2 &= |\alpha|^4 + 2 |\alpha|^2 \sinh^2(r) + \sinh^4(r) \\
    \mathrm{Re} \{ \langle a a \rangle \} &= |\alpha|^2 - \cosh(r) \sinh(r) \cos(\theta) \\
    \partial_\theta |\langle a a \rangle|^2 &= 2 |\alpha|^2 \cosh(r) \sinh(r) \sin(\theta) \\
    \partial_\theta \mathrm{Re} \{ \langle a a \rangle \} &= \cosh(r) \sinh(r) \sin(\theta)
\end{align*}

\section{Proof inequality for multi-mode fields}

In this section, we show that the quantum non-Gaussianity bound~(18) also holds for Gaussian multi-mode fields. We show it first for pure states. The generalization to mixed states then follows from the proof for mixed states given in the main text. The total photon number operator of $M$ modes is given by $n_\mathrm{tot} = \sum_i^M a^\dagger_i a_i$. According to the Bloch-Messiah reduction, any M-mode Gaussian pure state can be created by combining $M$ single-mode squeezed states using passive linear optical elements (beam splitters and phase shifters) \cite{braunstein_squeezing_2005, jezek_experimental_2011, weedbrook_gaussian_2012, cariolaro_bloch-messiah_2016, walschaers_nongaussian_2021}. 
Crucially, the corresponding unitary $U_\mathrm{L}$ of the linear optical elements does not change the total photon number operator $n_\mathrm{tot} = U_\mathrm{L} n_\mathrm{tot} U_\mathrm {L}^\dagger$. This means that the multi-mode correlation functions are also invariant under $U_\mathrm{L}$, since 
\begin{equation}
    g^{(2)}_\mathrm{tot} = \frac{\langle  n_\mathrm{tot} ( n_\mathrm{tot} - 1 )   \rangle }{ \langle n_\mathrm{tot} \rangle^2 } 
\hspace{1cm} \mathrm{and} \hspace{1cm} 
g^{(3)}_\mathrm{tot} = \frac{\langle  n_\mathrm{tot} ( n_\mathrm{tot} - 1 ) ( n_\mathrm{tot} - 2 )   \rangle }{ \langle n_\mathrm{tot} \rangle^3 }.
\label{eq:g2_g3_tot}
\end{equation}
Hence, reversing the statement of the Bloch-Messiah reduction means, we can always find a passive linear transformation that leads to a state with $M$ uncorrelated single-mode squeezed states, with unchanged measurement results for the correlation functions. This allows us to prove the inequality for multiple uncorrelated single-mode fields rather than a Gaussian multi-mode field without loss of generality. 

We use the inequality~(15) where we move the last term to the right-hand side and then square it, which gives 
\begin{equation}
    G^{(3)} + 9  G^{(2)} G^{(1)} + 6 \sqrt{ G^{(1)} G^{(2)} G^{(3)}} \geq 4 \Big[ G^{(1)} \Big]^3. 
    \label{eq:ineq_mm}
\end{equation} 
For a field with $M$ uncorrelated modes, the $n$-th order correlation functions become 
\begin{align}
    G^{(1)}_\mathrm{mm} &= \sum_i^M G_i^{(1)} \\
    G^{(2)}_\mathrm{mm} &= \sum_i^M G_i^{(2)} + 2 \sum_i^M \sum_{j>i}^M G_i^{(1)} G_j^{(1)} \\
    G^{(3)}_\mathrm{mm} &= \sum_i^M G_i^{(3)} + 3 \sum_i^M \sum_{j>i}^M \Bigg[ G_i^{(2)} G_j^{(1)} + G_i^{(1)} G_j^{(2)} \Bigg] +  6 \sum_i^M \sum_{j>i}^M \sum_{k>j}^M G^{(1)}_i G^{(1)}_j G^{(1)}_k
\end{align}
Note that the above definition ensures $g^{(2)}$ and $g^{(3)}$ values of unity for multiple modes of coherent fields. In the following, we insert these expressions $G^{(n)} \rightarrow G^{(n)}_\mathrm{mm}$ in the inequality~\eqref{eq:ineq_mm} to verify the validity of the bound for multi-mode fields. We obtain 
\begin{align}
    &\sum_i^M G_i^{(3)} 
    + 6 \sum_i^M \sum_{j>i}^M G_i^{(2)} G_j^{(1)} 
    + 6 \sum_i^M \sum_{j>i}^M \sum_{k>j}^M G^{(1)}_i G^{(1)}_j G^{(1)}_k 
    + 9 \Bigg[ \sum_i^M G_i^{(2)} + 2 \sum_i^M \sum_{j>i}^M G_i^{(1)} G_j^{(1)} \Bigg] \sum_k^M G_k^{(1)} 
    + 6 \sqrt{ G^{(1)}_\mathrm{mm} G^{(2)}_\mathrm{mm} G^{(3)}_\mathrm{mm} } \nonumber \\
    & \, {\stackrel{?} \geq} \,  4 \sum_i^M \Big[ G_i^{(1)} \Big]^3 
    + 12 \sum_i^{M} \sum_{j>i}^{M} \Bigg( \Big[ G_i^{(1)} \Big]^2 G_j + G_i^{(1)} \Big[ G_j^{(1)} \Big]^2  \Bigg) 
    + 24 \sum_{i}^{M} \sum_{j>i}^{M} \sum_{k>j}^{M} G_i^{(1)} G_j^{(1)} G_k^{(1)}. \nonumber
\end{align}
Here, the last term on the right-hand side is compensated by the expression
\begin{equation*}
    6 \sum_i^M \sum_{j>i}^M \sum_{k>j}^M G^{(1)}_i G^{(1)}_j G^{(1)}_k 
    + 9 \cdot 2 \sum_i^M \sum_{j>i}^M G_i^{(1)} G_j^{(1)} \sum_k^M G_k^{(1)},
\end{equation*}
on the left-hand side. For this, the terms with $k=i$ and $k=j$ in the second term are not used. Therefore, these can compensate for the second term on the right-hand side. The remaining part to verify the inequality is to show that 
\begin{align}
    &\sum_i^M G_i^{(3)} + 9 \sum_i^M G_i^{(2)} \sum_k^M G_k^{(1)}  + 6 \sqrt{ G^{(1)}_\mathrm{mm} G^{(2)}_\mathrm{mm} G^{(3)}_\mathrm{mm} } \, {\stackrel{?} \geq} \,  4 \sum_i^M \Big[ G_i^{(1)} \Big]^3.
\end{align}
We further approximate the left-hand side by using $G^{(n)}_\mathrm{mm} \geq \sum_i^M G^{(n)}_i$ and $\sum_i^M G^{(n)}_i \sum_j^M G^{(m)}_j \geq \sum_i^M G^{(n)}_i  G^{(m)}_i $, which leads to
\begin{align}
    &\sum_i^M G_i^{(3)} + 9 \sum_i^M G_i^{(2)} G_i^{(1)}  + 6 \sqrt{  \sum_i^M G^{(1)}_i \sum_j^M G^{(2)}_j G^{(3)}_j } \, {\stackrel{?} \geq} \,  4 \sum_i^M \Big[ G_i^{(1)} \Big]^3.
    \label{eq:ineq_mm_2}
\end{align}
Utilizing the Cauchy-Schwarz inequality~\cite{cauchy1821formules} to derive the bound  
\begin{equation}
    \sqrt{  \sum_i^M G^{(1)}_i \sum_j^M G^{(2)}_j G^{(3)}_j } \geq \sum_i^M G_i^{(1)} G_i^{(2)} G_i^{(3)}
\end{equation}
proofs that the quantum non-Gaussianity bound~(18) also holds for Gaussian multi-mode fields since the inequality~\eqref{eq:ineq_mm_2} is valid for each of the $M$ modes, according to Eq.~\eqref{eq:ineq_mm}. The subsequent proof for mixed states in the main text is independent of whether the state is single or multi-mode. It is thus also valid for multi-mode fields.

\section{Linear bound - tangent lines} \label{sec:liner_bound}

\begin{figure}[h]
    \centering
    \includegraphics[width=0.4\textwidth]{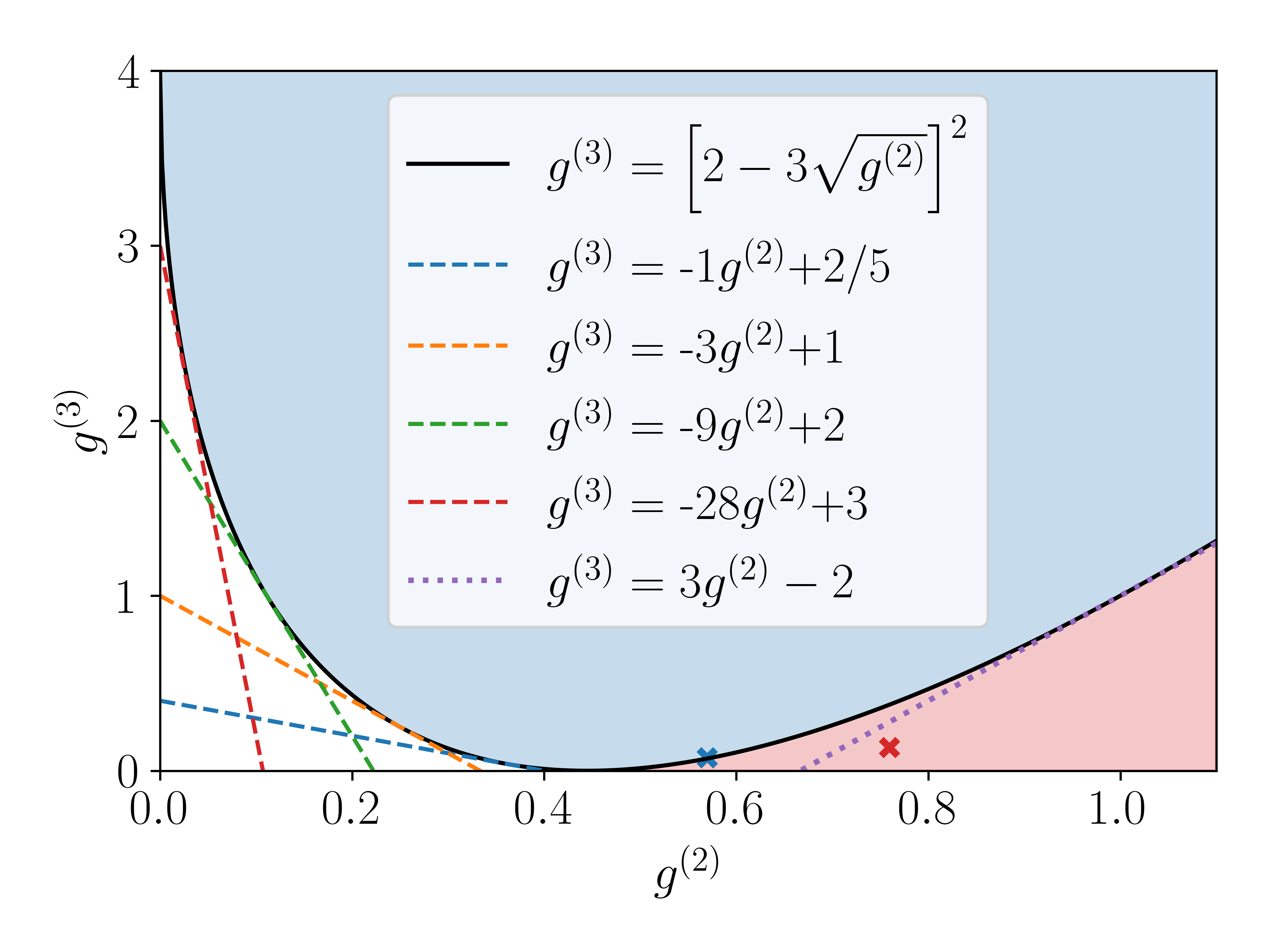}
    \caption{Linear bound - tangent lines. The dashed and dotted lines represent different (simpler) versions of the general bound, see Eq.~\eqref{eq:tangent_lines} and Eq.~\eqref{eq:g3_cumulant}.}
    \label{fig:g2g3_tangent_supp}
\end{figure}

The general bound for Gaussian pure states [Eq.~(14)] can be expressed with several, initially assumed, linear bounds [Eq.~(5)] by the slope of a tangent line 
\begin{align}
    k = - \frac{3 \Big( 2 - 3 \sqrt{g^{(2)}} \big)}{ \sqrt{g^{(2)}} },
    \label{eq:tangent_slope}
\end{align}  
corresponding to $\chi_2 = -k$ in Eq.~(5). The associate intercept is 
\begin{align}
    \chi_1 = 2 + 3\sqrt{g^{(2)}} - 9 g^{(2)}.
    \label{eq:tangent_intercept}
\end{align}  
Note that the maximal slope is $k = 3$, which can be seen in Eq.~\eqref{eq:tangent_slope}, and hence the requirement $\chi_2 \geq -3$ is satisfied. Measuring a combination of $g^{(2)}$ and $g^{(3)}$, which lies below one of the following tangent lines (see Eq.~(5) and Fig.~\ref{fig:g2g3_tangent_supp}) also serves as an unambiguous proof of quantum non-Gaussianity:
\begin{subequations}
\begin{eqnarray}
    & g^{(3)} + g^{(2)} < 2/5 \\
    & g^{(3)} + 3 g^{(2)} < 1 \\
    & g^{(3)} + 9 g^{(2)} < 2 \\
    & g^{(3)} + 28 g^{(2)} < 3
\end{eqnarray}
\label{eq:tangent_lines}
\end{subequations}
These four inequalities already cover most of the (white) quantum non-Gaussianity region in Fig.~\ref{fig:g2g3_tangent_supp}.

For Gaussian pure states, a simpler bound, corresponding to a tangent line, is given by the joint cumulant (connected component)~\cite{kubo1962generalized, stiesdal2018observation, plankensteiner2021quantumcumulantsjl} of the third-order correlation function 
\begin{align}
    g^{(3)}_c = g^{(3)} - 3 g^{(2)} + 2,
    \label{eq:g3_cumulant}
\end{align}
which needs to be positive for Gaussian pure states. This quantity also has a meaningful interpretation as the ``pure" three-particle contributions of $g^{(3)}$. This can be seen more intuitively by writing the joint cumulant of a third-order term as $ \langle {:} n_1 n_2 n_3 {:} \rangle_c = \langle {:} \Delta n_1 \Delta n_2 \Delta n_3 {:} \rangle$, with $\Delta n_i = n_i - \langle n_i \rangle$. Note, however, that this is only valid up to third-order since above the third order the joint cumulant does not correspond to the central moment for higher $n$~\cite{kubo1962generalized}. 

\section{Three-fold coincidences}
\begin{figure}[h]
    \centering
    \includegraphics[width=\linewidth]{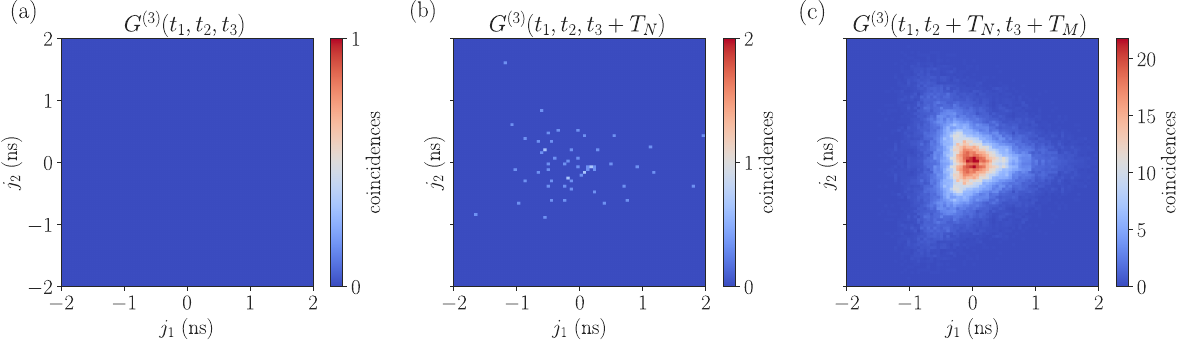}
    \caption{\textbf{Measured three-fold coincidences.} The detected coincidences for (a) the same pulse, (b) a pair of subsequent pulses, and (c) three subsequent pulses, each after transformation into Jacobi coordinates $j_1 = (2t_1 - t_2 - t_3)/\sqrt{6}$ and $j_2 = (t_2-t_3)/\sqrt{2}$ with $t_i$ the time at detector $i$. 
    }
    \label{fig:g3_full_FB}
\end{figure}

Measured three-fold coincidences are shown in \fref{fig:g3_full_FB}. 
As discussed in the main text, zero three-photon events are detected for the same excitation pulse, see \fref{fig:g3_full_FB}(a). 
For a pair of subsequent pulses, where we condition on two clicks in one pulse and one photon in the other, see \fref{fig:g3_full_FB}(b), there are only a few three-photon events, which is consistent with the low value of $g^{(2)}$. 
In contrast, there are ample three-photon events for three separate pulses as shown in \fref{fig:g3_full_FB}(c). The coincidences form a threefold rotationally symmetric ($120^\circ$) pattern, due to the equivalence of the three time coordinates and the exponential profile of the photons emitted by the QD. A time-symmetric temporal profile would feature a six-fold rotational symmetry ($60^\circ$).

Note that (b) and (c) are both combinations of six different temporal orderings. These orderings are based on which of the detectors clicked for which of the laser pulses. To give an example, an event in (b) would be that detector 1 clicked due to a certain excitation pulse, and detectors 2 and 3 clicked due to the subsequent pulse. More details on the analysis for the three-fold coincidences are provided in Ref.~\cite{Hansen2026}.

\begin{figure}[h]
    \centering
    \includegraphics[width=\linewidth]{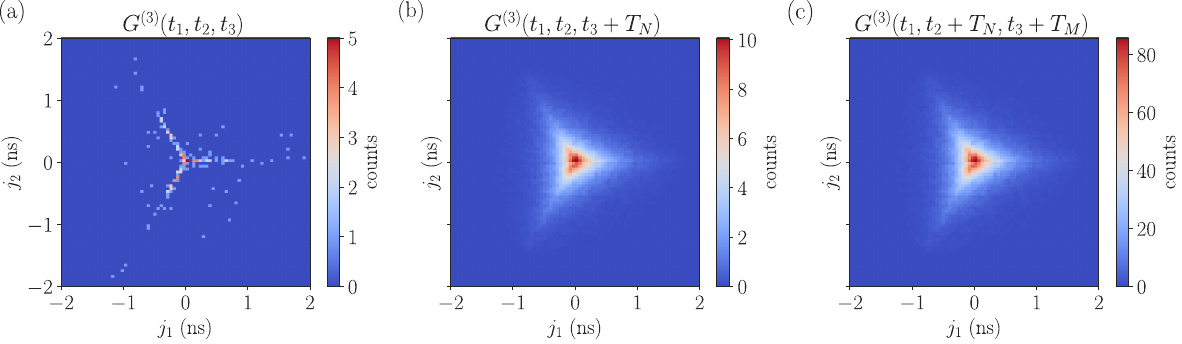}
    \caption{\textbf{Measured three-fold coincidences with increased laser leakage.} The detected coincidences for (a) the same pulse, (b) a pair of subsequent pulses, and (c) three subsequent pulses, each after transformation into Jacobi coordinates. }
    \label{fig:g3_full_noFB}
\end{figure}

In a separate experimental run, as a check for our setup and analysis, threefold coincidences are measured with intentionally increased laser leakage.
The leakage of laser light into the waveguide is increased by misaligning the optical path for the excitation.
In addition, no spectral filtering is used for the collection, thus a larger portion of the spectrally broader laser light propagates to the detectors. 
Despite the increased laser light contribution, the signal is still dominated by emission from the quantum dot. 
Figure~\ref{fig:g3_full_noFB} shows the measured three-fold coincidences with increased laser leakage. 
In this case, substantially more three-photon events are detected for the same excitation pulse, see \fref{fig:g3_full_noFB}(a). 
As mentioned in the main text, the temporally narrow features are caused by the fact that two detections correspond to laser photons, which are temporally narrow and at e.g.\ $t_1 \approx t_2 \approx 0$, while the third detection is from QD emission, which has an exponential decay profile. 
For \fref{fig:g3_full_noFB}(b) the statistics are increased as well in comparison to \fref{fig:g3_full_FB}(b). In this case, the temporal profile is a consequence of one photon coming from the laser and two photons from QD emission after separate laser pulses. 
The difference in temporal profile in \fref{fig:g3_full_FB}(c) and \fref{fig:g3_full_noFB}(c) is explained by a reduced detection jitter for the latter. This does however not affect the acquired number of coincidences. 

\section{P-value test}
To perform a p-value test, three essential components are required. First, a null hypothesis, representing the model or assumption to be tested. Second, a probability distribution describing the expected statistical fluctuations of the relevant observables under the null hypothesis; this enables the calculation of the likelihood of any particular outcome. Third, a test statistic or observed quantity must be identified, which can be evaluated against the predicted distribution.

The p-value is then defined as the cumulative probability of obtaining outcomes at least as extreme as the observed measurement. A small p-value indicates that the observed data lie in the tail of the distribution predicted by the null hypothesis, thereby providing statistically significant evidence to reject it.

In our case, the null hypothesis is that the measured state is Gaussian and its correlations lie on the boundary. Assuming a fixed number of single detections $N_1$ and some value of $g^{2}$ and $g^{(3)}$, the  expected number of 2- and 3-photon events can be calculated as
$N_{2,0}=\frac{g^{(2)}\cdot N_1^2}{N_{shots}}$ and $N_{3,0}=\frac{g^{(3)}\cdot N_1^3}{N_{shots}^2}$, where $N_{shots}=2.963 \cdot 10^{11}$ is the total number of excitation pulses and $N_1=7.041\cdot 10^8$ is the number of single-photon events. For example, for a particular point on the boundary $g^{(2)}=0.2$ and $g^{(3)}=0.43$ we obtain $N_{2,0}=3.346\cdot 10^{5}$ and $N_{3,0}=1723$. $N_{2,0}$ and $N_{3,0}$ are the number of 2-photon and 3-photon events expected under the assumption that the null-hypothesis is true. 

Assuming Poissonian photon counting statistics, the joint probability of observing specific count events $N_{2}$ and $N_{3}$ is P$(N_{2},N_{3})=\text{Pois}(N_{2}|N_{2,0})\cdot \text{Pois}(N_{3}|N_{3,0})$. For the measured 2- and 3-photon counts $N_{2,m}=1.96\cdot 10^4$ and $N_{3,m}=0$ we find P$(N_{2,m},N_{3,m})=e^{-2.61\cdot 10^{5}}=1\cdot 10^{-113351}$. 
The combined p-value is defined as $\tilde{p}= \sum_{P(N_2,N_3)<P(N_{2,m},N_{3,m})} \text{P}(N_2, N_3)$. For $N_{2,m}=1.96\cdot 10^4$ and $N_{3,m}=0$ we find $\tilde{p}= e^{-2.61\cdot 10^{5}} = 1.2\cdot 10^{-113402}$. Maximizing $\tilde{p}$ over all values of $g^{(2)}$ and $g^{(3)}$ at the bound yields a final upper bound of $p=4\cdot 10^{-4793}$. This extremely small p-value indicates that the measured data is highly inconsistent with the Gaussian-state hypothesis.

\section{Multi-photon contribution and comparison to an existing criterion} 
\begin{figure}[h]
    \centering
    \includegraphics[width=0.99\textwidth]{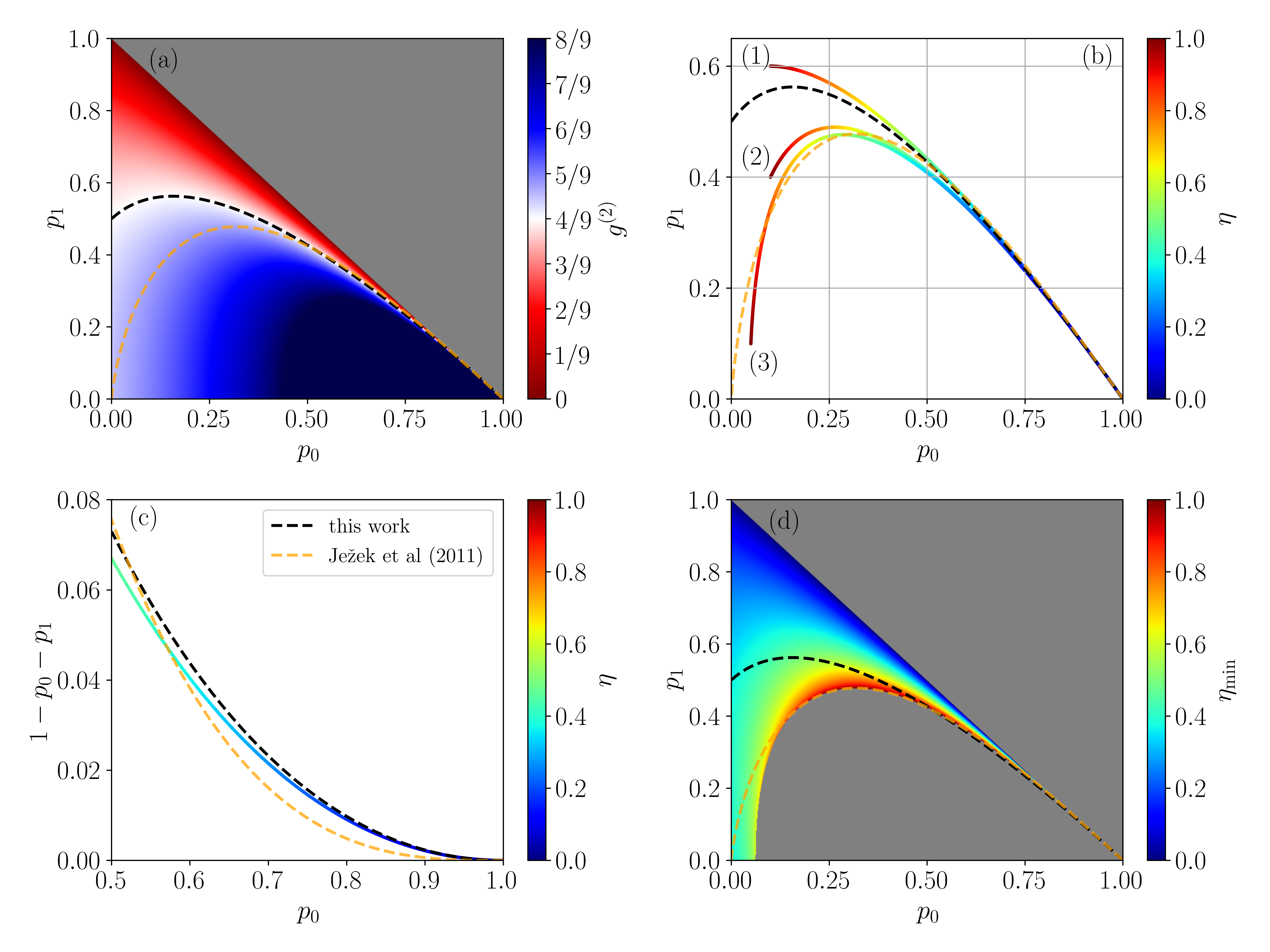}
    \caption{\textbf{Multi-photon contribution.}  (a) $g^{(2)}$ for the multi-photon state $\rho$ of Eq.~\eqref{eq:ph_state}. Since for this particular $\rho$ we always have $g^{(3)}=0$, the developed criterion requires $g^{(2)} < 4/9$ for QNG. In all plots, the black dashed line shows $g^{(2)} = 4/9$ and the orange dashed line shows the bound derived in Ref.~\cite{jezek_experimental_2011}. Around $p_0 \approx 1/2$ the two lines cross, see also plot (c). The gray area is not physical. (b) Trajectories where the efficiency $\eta$ (loss $1-\eta$) is varied from unity to zero, for three different initial states with $(p_0,p_1) = \{ (0.1,0.5), \, (0.1,0.4), \, (0.05, 0.1)  \}$. (c) Two-photon population bound for $p_0 > 1/2$, which serves as a "zoom-in" of the crossing of the bounds. Note that here the state needs to be below the line to be quantum non-Gaussian. (d) Minimal efficiency $\eta_\mathrm{min}$ (maximal loss $1-\eta_\mathrm{min}$) required to detect the QNG of a state with the criterion derived in Ref.~\cite{jezek_experimental_2011}. 
    } 
    \label{fig:comp_exs}
\end{figure}

The developed criterion is not capable of detecting the quantum non-Gaussianity (QNG) of Fock-states with $n \geq 2$. For a single-photon source, this means that multi-photon contributions can bring the state out of the detectable region. In the following, we investigate the robustness to two-photon contributions for a density matrix of the form 
\begin{equation}
    \rho = p_0 | 0 \rangle \langle 0 | +  p_1 | 1 \rangle \langle 1 | +  (1 - p_0 - p_1) | 2 \rangle \langle 2 |.
    \label{eq:ph_state}
\end{equation}
The same analysis also holds for a state $|\psi\rangle = e^{i\phi_0} \sqrt{p_0} | 0 \rangle + e^{i\phi_1} \sqrt{p_1} | 1 \rangle + e^{i\phi_2} \sqrt{1-p_0-p_1} | 2 \rangle$ with arbitrary phases $\phi_i$. Since for this state we have $g^{(3)} = 0$, $g^{(2)}$ needs to be smaller than $4/9$ for our criterion to detect QNG. In Fig.~\ref{fig:comp_exs}(a), we show $g^{(2)}$ for the above state with a scan of $p_0$ and $p_1$. For the red region, $g^{(2)}$ is smaller than $4/9$, which means QNG can be detected. The black dashed line shows the boundary to $g^{(2)} > 4/9$, which is given by 
\begin{equation}
    p_1 = \frac{1}{4}( 3\sqrt{8p_0 + 1} - 1 - 8p_0 ),
    \label{eq:p1_p0}
\end{equation}
corresponding to $g^{(2)} = 2(1-p_0-p1)/(2-2p_0-p_1)^2 \leq 4/9$. We can see that a considerable amount of two-photon contribution $(1-p_0-p_1)$ is tolerable for our criterion. For comparison, the orange dashed line shows the non-Gaussianity witness derived in Ref.~\cite{jezek_experimental_2011}. 
With perfect detection efficiency and no loss, this criterion is able to verify QNG for additional states if $p_0 \lesssim 1/2$ (region between the black and orange dashed lines); however, above $p_0 \gtrsim 1/2$, the two bounds cross, which allows us to verify the QNG for more states with the current criterion. The crossing can be seen more clearly in Fig.~\ref{fig:comp_exs}(c). This region is particularly important for attenuated states. Since the witness of Ref.~\cite{jezek_experimental_2011} is not loss-resistant, there is a minimum efficiency $\eta$ (including detection and loss) for the state to be detectable as quantum non-Gaussian. Loss with a probability of $1-\eta$ changes a mixed state $\sum_k^\infty p_{k=0} | k \rangle \langle k |$ to
\begin{equation}
    \rho_{\eta} = \sum_{k=0}^{\infty} \left( \sum_{n=k}^{\infty} p_n \binom{n}{k} \eta^k (1-\eta)^{n-k} \right) \lvert k \rangle \langle k \rvert. 
    \label{eq:state_loss}
\end{equation}
For our initial state $\rho$ in Eq.~\eqref{eq:ph_state}, this means we obtain $\rho_{\eta} = P_0 \lvert 0 \rangle \langle 0 \rvert + P_1 \lvert 1 \rangle \langle 1 \rvert + P_2 \lvert 2 \rangle \langle 2 \rvert$, with $P_0 = p_0 + p_1 (1 - \eta) + p_2 (1 - \eta)^2$, $P_1 = p_1 \eta + 2 p_2 \eta (1 - \eta)$ and $P_2 = p_2 \eta^2$. 

In Fig.~\ref{fig:comp_exs}(b), we show the evolution of three different states with an efficiency $\eta$ ranging from unity to zero. For evolution (1), we can see that it crosses the orange line at around $\eta \approx 0.4$. To see this more clearly, we plot the same curves in Fig.~\ref{fig:comp_exs}(c) but with $1-p_0-p_1$ on the y-axis, which corresponds to the two-photon component, and equals the vertical distance to the non-physical region in (a). Note that the $\eta$-evolution does not cross the black line, originating from the loss-resistance of our criterion. 
The efficiency $\eta$ at which the evolution crosses the yellow line, as described above, determines a minimal efficiency $\eta_\mathrm{min}$ to detect QNG with the criterion in Ref.~\cite{jezek_experimental_2011}. We plot $\eta_\mathrm{min}$ in Fig.~\ref{fig:comp_exs}(d). The QNG of states above the black dashed line, which cannot be measured with an efficiency above $\eta_\mathrm{min}$, can always be verified by our criterion, but not by the other, if the efficiency is below $\eta_\mathrm{min}$. 
On the other hand, non-Gaussianity of states below the dashed black line but above the orange line cannot be verified with our criterion but with the one derived in Ref.~\cite{jezek_experimental_2011} if $\eta > \eta_\mathrm{min}$, see also evolution (2) in Fig.~\ref{fig:comp_exs}(b). 
Finally, we also identify an initially surprising region below the orange line in Fig.~\ref{fig:comp_exs}(d), with a finite minimal efficiency $\eta_\mathrm{min}$ even though this corresponds to states which cannot be detected with the criterion in Ref.~\cite{jezek_experimental_2011} for $\eta=1$. This can be explained with evolution (3) in Fig.~\ref{fig:comp_exs}(b): Additional losses can shift a state into the detectable QNG region due to the population increase of the single photon component from a two photon state. The plotted $\eta_\mathrm{min}$ in Fig.~\ref{fig:comp_exs}(d) in this region defines when the orange line is crossed again Fig.~\ref{fig:comp_exs}(b). 

\section{Criterion based on $g^{(2)}$ and the mean photon number $\langle a^\dagger a \rangle$}

During the derivation of the QNG criterion based on the second and third-order correlation functions $g^{(2)}$ and $g^{(3)}$, we additionally found a criterion based on the mean photon number $\langle a^\dagger a \rangle = G^{(1)}$ and $g^{(2)}$. The same criterion has also been independently derived in Refs.~\cite{kalash_certifying_2025, racz_witnessing_2025}. The photon number detection is, however, not resistant to losses. This means, certifying the QNG with this criterion requires high detection efficiency. If the mean photon number and $g^{(2)}$ are measured for sources with high efficiency, the procedures typically employed for measuring these quantities in the low efficiency regime are not necessarily applicable, and photon number resolving detectors should preferably be used.  
For $p_i$ being the probability to have a Fock state $| i \rangle$, a photon number resolving detector measures the quantity $\langle a^\dagger a \rangle_{\mathrm{PNR}} = \sum_i i p_i$, whereas a non-photon number resolving detector measures the probability $\langle a^\dagger a \rangle_{\mathrm{NPNR}} = \sum_i p_i$. These quantities are not the same unless we are dominated by the single photon probability $p_1$, which happens automatically for low efficiency sources, but may not be the case for sources with high efficiency. Hence, one needs to be careful that the measured quantity corresponds to the one entering the criterion. 

\begin{figure}[h]
    \centering
    \includegraphics[width=0.5\textwidth]{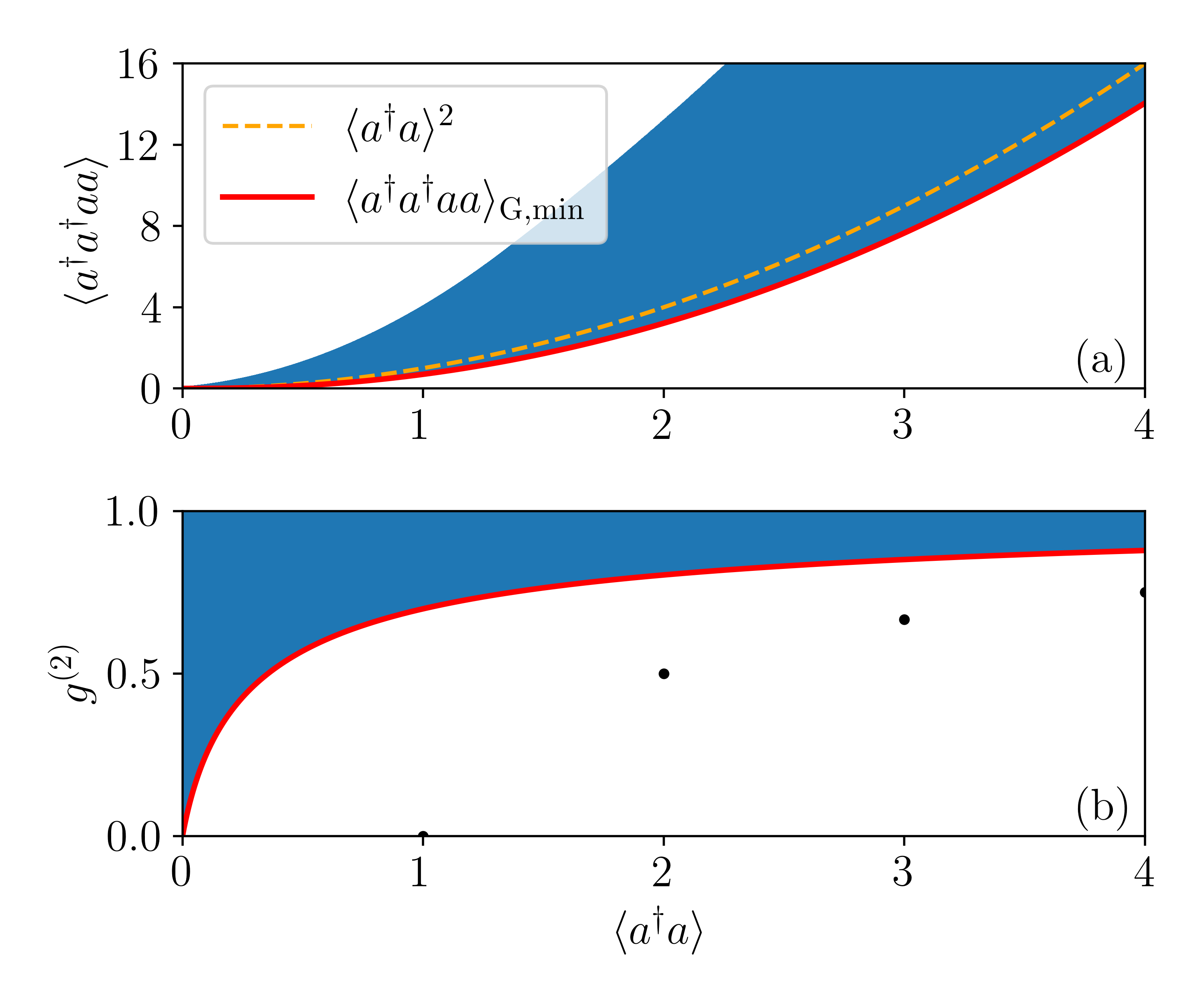}
    \caption{Bound based on the mean photon number and two photon correlations. 
    The blue points represent a large set of combinations of $\alpha$, $r$ and $\theta$. The red solid line depicts the analytic bound for Gaussian states described by Eq.~\eqref{eq:G2_min}. (a) shows the bound for $G^{(2)} = \langle a^\dagger a^\dagger a a \rangle$ depending on $G^{(1)} = \langle a^\dagger a \rangle$ and (b) depicts the equivalent threshold for $g^{(2)} = G^{(2)}/[G^{(1)}]^2$. The black dots represent the first four Fock states. The dashed orange line in (a) corresponds to coherent states ($g^{(2)} = 1$). For the scatter plot we use 1000 equally distributed values in the interval $\alpha \in (0, 2]$, 501 values for $r \in [0, 2]$ and 21 values for $\theta \in [0, \pi ]$.} 
    \label{fig:G2G1}
\end{figure}

In Fig.~\ref{fig:G2G1} we plot $\langle a^\dagger a \rangle$ and $\langle a^\dagger a^\dagger a a \rangle$ for different Gaussian states (blue region), which indicates that they form a convex surface. 
In the following, we determine the minimum value of $\langle a^\dagger a^\dagger a a \rangle$ corresponding to a certain mean photon number $\langle a^\dagger a \rangle$. We start by calculating the first and second derivative of $G^{(2)}_\mathrm{G} = \langle a^\dagger a^\dagger a a \rangle_\mathrm{G}$ with respect to $\theta$
\begin{align}
    & \partial_\theta G^{(2)}_\mathrm{G} = 2 |\alpha|^2 \cosh(r) \sinh(r) \sin(\theta) \\
    & \partial^2_\theta G^{(2)}_\mathrm{G} = 2 |\alpha|^2 \cosh(r) \sinh(r) \cos(\theta). 
\end{align}
This shows that the minimum of $G^{(2)}_\mathrm{G}$ is always given for $\theta = 0$, regardless of the photon number $\langle a^\dagger a \rangle$. 
Besides this, both quantities depend on $\alpha^2$ and $r$ [see Eq.~(4)], therefore we will use the Lagrange multiplier method to determine the minimum. This can be done by minimizing the function 
\begin{align}
    \mathcal{L}(\alpha^2,r,\lambda) = \langle a^\dagger a^\dagger a a \rangle(\alpha^2, r) + \lambda \Big[  \langle a^\dagger a \rangle(\alpha^2, r) - n_c \Big] ,
    \label{eq:lagrange}
\end{align}
where $n_c$ defines the constant photon number $\langle a^\dagger a \rangle$ for which the corresponding minimal $\langle a^\dagger a^\dagger a a \rangle_\mathrm{G, min}$ for Gaussian states is calculated. 
We follow the standard procedure and calculate the partial derivatives for all variables and set them to zero:
\begin{subequations}
\begin{eqnarray}
    & \partial_\lambda \mathcal{L} = \langle a^\dagger a \rangle(\alpha^2, r) - n_c = 0 \\
    & \partial_{\alpha^2} \mathcal{L} = 4 \sinh^4(r) - 2 \sinh(r) \cosh(r) + 2 \alpha^2 + \lambda = 0 \\
    & \partial_{r} \mathcal{L} = 2 \Big[ \sinh(r) \cosh(r) ( \lambda 5 \sinh^2(r) + \cosh^2(r) ) - \alpha^2 ( \sinh^2(r) + \cosh^2(r) -4 \sinh(r) \cosh(r) ) \Big] = 0
\end{eqnarray}
\label{eq:lagrange_deriv}
\end{subequations}
Inserting the first two equations in the last one to eliminate $\alpha^2$ and $\lambda$, and some algebra leads to
\begin{align}
    -(4 n_c + 2) + e^{-2r} + e^{6r} = 0
    \label{eq:lagrange_deriv_2}
\end{align}
With the substitution $x = \exp (2r)$ we obtain the quartic equation 
\begin{align}
    1 + x^4 - (4 n_c + 2) x = 0
    \label{eq:quardic}
\end{align}
with the solution
\begin{align}
    x( \langle a^\dagger a \rangle ) &= \frac{1}{2} \sqrt{ \frac{C( \langle a^\dagger a \rangle )}{\sqrt[3]{18}} + \frac{4 \sqrt[3]{2/3}}{C( \langle a^\dagger a \rangle )}} + \frac{1}{2}  \sqrt{ \frac{8 \langle a^\dagger a \rangle + 4}{ \sqrt{ \frac{C( \langle a^\dagger a \rangle )}{\sqrt[3]{18}} + \frac{4 \sqrt[3]{2/3}}{C( \langle a^\dagger a \rangle )} } }  - \left[ \frac{C( \langle a^\dagger a \rangle )}{\sqrt[3]{18}} + \frac{4 \sqrt[3]{2/3}}{C( \langle a^\dagger a \rangle )} \right]}
    \label{eq:quardic_sol}
\end{align}
where $C( \langle a^\dagger a \rangle ) = \sqrt[3]{ \sqrt{3} \sqrt{27(4  \langle a^\dagger a \rangle + 2)^4 - 256} + 9(4  \langle a^\dagger a \rangle + 2)^2}$ and we have inserted $n_c = \langle a^\dagger a \rangle$. 
The minimum value for $ \langle a^\dagger a^\dagger a a \rangle $ with respect to $ \langle a^\dagger a \rangle$ is then given by $r = ln(x) / 2$ and the corresponding $\alpha^2 =  \langle a^\dagger a \rangle  - \sinh^2(r)$. We verify numerically that this gives a minimum and not a maximum or saddle point. 
Inserting this in the expression for $\langle a^\dagger a^\dagger a a \rangle $ leads to 
\begin{align}
    \langle a^\dagger a^\dagger a a \rangle_\mathrm{G, min} = \frac{x^4 + x^2 (8 \langle a^\dagger a \rangle^2 - 8  \langle a^\dagger a \rangle - 4) + x (8  \langle a^\dagger a \rangle +4) - 1 }{8 x^2} ,
    \label{eq:G2_min}
\end{align}
with $x$ given by Eq.~\eqref{eq:quardic_sol}. 

The red line in Fig.~\ref{fig:G2G1}(a) depicts Eq.~\eqref{eq:G2_min}. We see that it defines the boundary of the Gaussian states. Furthermore, due to the linearity of expectation values, a mixture of states directly translates to the expectation values, i.e.\ $\langle \mathcal{O} \rangle_\rho = \sum_i p_i \langle \mathcal{O} \rangle_{\psi_i}$ with $\rho = \sum p_i | \psi_i \rangle \langle \psi_i |$. This means that a mixture of two points in Fig.~\ref{fig:G2G1}(a) is always on the connecting straight line between them. Since the boundary for the Gaussian pure states describes a convex set, the non-Gaussianity criterion is directly extended to mixed states and multi-mode fields. We note that this convexity argument applies to expectation values. The non-Gaussian area in Fig.~1 of the main text is also convex, but the same convexity argument cannot be applied  since $g^{(2)}$ and $g^{(3)}$ are not expectation values. 

Dividing $\langle a^\dagger a^\dagger a a \rangle_\mathrm{G, min}$ by the corresponding $\langle a^\dagger a \rangle^2$ leads to the mean photon number dependent threshold for the correlation function $g^{(2)}_\mathrm{G,min}$, shown in Fig.~\ref{fig:G2G1}(b). For large mean photon numbers, the non-Gaussianity bound approaches $g^{(2)}_\mathrm{G, min} \rightarrow 1$. We numerically tested the bound for Fock states $| n \rangle$ up to $n = 1000$, which are all below the bound. This means that this bound could, in principle, be used to verify them. The black dots in Fig.~\ref{fig:G2G1}(b) represent the first four Fock states. 

\end{document}